\documentclass[conference]{IEEEtran}
\IEEEoverridecommandlockouts
\usepackage{booktabs}
\usepackage{multirow}
\usepackage{textcomp}
\usepackage{xcolor}
\usepackage{cite}
\usepackage{latexsym}
\usepackage{graphicx}
\usepackage{amsfonts,amssymb,amsmath}
\usepackage{nccmath}
\usepackage{optidef}
\usepackage{hyperref}
\usepackage{color,soul}
\usepackage{array}
\usepackage{algorithm}
\usepackage{algorithmic}
\usepackage[T1]{fontenc}
\usepackage[utf8]{inputenc}
\usepackage{fancyhdr}
\usepackage{lastpage}
\usepackage{caption}
\usepackage{subcaption}
\usepackage{setspace}

\def\BibTeX{{\rm B\kern-.05em{\sc i\kern-.025em b}\kern-.08em
    T\kern-.1667em\lower.7ex\hbox{E}\kern-.125emX}}

\usepackage{bm}

\fancypagestyle{firststyle}{
	\fancyhf{}
	\fancyhead[L]{H.Poddar, T. Yoshimura, M. Pagin, T. S. Rappaport, A. Ishii and M. Zorzi, ``Full-Stack End-To-End mmWave Simulations Using 3GPP and NYUSIM Channel Model in ns-3'' \textit{in ICC 2023 - 2023 IEEE International Conference on Communications,} Rome, Italy, May 2023, pp. 1--6.}

}

\usepackage{soul}
\begin{document}

\title{Full-Stack End-To-End mmWave Simulations Using 3GPP and NYUSIM Channel Model in ns-3
		\thanks{This work is supported by the NYU WIRELESS industrial affiliates program and the commissioned research (No.04201) from the National Institute of Information and Communications Technology (NICT), JAPAN.}}
	
	\author{\IEEEauthorblockN{Hitesh Poddar$^{*}$, Tomoki Yoshimura$^{\dagger}$, Matteo Pagin$^{\ddagger}$, Theodore S. Rappaport$^{*}$, Art Ishii$^{\dagger}$, and Michele Zorzi$^{\ddagger}$}
		\IEEEauthorblockA{$^{*}$NYU WIRELESS, NYU Tandon School of Engineering,\{hiteshp, tsr\}@nyu.edu\\
			$^{\dagger}$Corporate Research and Development Group, Sharp Corporation, Vancouver WA, USA,\{yoshimurat, ishiia\}@sharplabs.com\\
				$^{\ddagger}$SIGNET Lab, University of Padova, Italy,\{paginmatte, zorzi\}@dei.unipd.it}
	}

\maketitle
\thispagestyle{firststyle}
\linespread{1.05}
\begin{abstract}
Accurate channel modeling and simulation tools are vital for studying sub-THz and millimeter (mmWave) wideband communication system performance. To accurately design future high data rate, low latency wireless modems, the entire protocol stack must be appropriately modeled to understand how the physical layer impacts the end-to-end performance experienced by the end user. This paper presents a full stack end-to-end performance analysis in ns-3 using drop-based NYU channel model (NYUSIM)  and 3GPP statistical channel model (SCM) in scenarios, namely urban microcell (UMi), urban macrocell (UMa), rural macrocell (RMa), and indoor hotspot (InH) at 28 GHz with 100 MHz bandwidth. Video data is transmitted at 50 Mbps using User Datagram Protocol (UDP), and we observe that the RMa channel is benign in non-line of sight (NLOS) for NYUSIM and 3GPP SCM  as it exhibits no packet drops and yields maximum throughput (48.1 Mbps) and latency of $\sim$ 20 ms. In NLOS, for NYUSIM, the UMa and RMa channels are similar in terms of throughput and packet drops, and the latency in UMi and InH scenarios is 10 times and 25 times higher respectively compared to UMa.
Our results indicate that mmWave bands can support data rates of 50 Mbps with negligible packet drops and latency below 150 ms in all scenarios using NYUSIM.
\end{abstract}


\section{Introduction}
6G, or Beyond 5G, is expected to achieve larger capacity and lower latency through wider bandwidth channels in the mmWave and sub-THz spectrum \cite{singh2019beyond,NextGAlliance}. An accurate, statistical spatial and temporal channel model applicable for many different use cases in fixed and mobile applications (4K, 8K video streaming, AR/VR) is needed so that full-stack end-to-end network simulations may be conducted over a wide range of frequency spectrums. Discrete-event network simulators such as ns-3 are widely used by researchers worldwide to investigate and analyze complex wireless networks and design new protocols. The mmWave \cite{mezzavilla2018end} and New Radio (NR) \cite{patriciello2019e2e} modules in ns-3 are the most popular full-stack end-to-end simulation platforms for simulating 5G mmWave networks using the 3GPP SCM  \cite{zugno2020implementation},\cite{3GPPTR}. The simulation of a complex network with multiple user equipment (UEs) and base stations (gNBs), each featuring a detailed protocol stack implementation, is computationally intensive. Thus, SCMs are used for system-level simulations in ns-3 as they provide the best trade-off between accuracy and complexity. Among the existing channel models \cite{Rap2017itap}, the ones most used by academia and industry for simulating 5G wireless systems are the TR 38.901 (release 16) and NYUSIM, developed by 3GPP and NYU WIRELESS, respectively \cite{3GPPTR},\cite{sun2017novel},\cite{ju2019millimeter}. Prior work \cite{rappaport2017investigation} has shown that NYUSIM offers a more realistic characterization of the wireless channel than 3GPP SCM  in outdoor scenarios. Using hybrid precoding and analog combining method proposed in \cite{7160780}
and the digital block diagonalization (BD) approach presented in \cite{1261332} for a single-cell three-user MIMO scenario, the spectral efficiency (SE) per user (averaged over three users) with one data stream and one RF chain per user generated by NYUSIM and 3GPP SCM  was $\sim$ 18 bits/Hz and 12 bits/Hz in \cite{rappaport2017investigation}.\\ An increase in SE by 1.5 times in NYUSIM compared to 3GPP SCM  also implies a higher SINR in NYUSIM compared to 3GPP SCM. This motivated us to implement NYUSIM in ns-3 so that researchers can understand the impact of the lower-layers discrepancies such as SE found in \cite{rappaport2017investigation, alizadeh2019study,sun2018propagation} on the end-to-end throughput, latency and packet drop of a wireless modem.
Our implementation of NYUSIM in ns-3 is based on the identical open-source mmWave and sub-THz wireless channel simulator \textit{NYUSIM}. \textit{NYUSIM} \cite{sun2017novel},\cite{ju2019millimeter} was created using extensive field measurements over 2 terabytes of measurement data from 28 GHz to 140 GHz obtained during 2011-2022 in various scenarios such as UMi, UMa, RMa, InH, and Indoor Factory (InF) \cite{rappaport2013millimeter,rappaport2015wideband,samimi20163,samimi2016local,sun2016investigation,sun2015synthesizing,maccartney2015indoor,maccartney2016millimeter,maccartney2017rural,ju20203,ju2021millimeter,ju2022sub,ju2022Submitted}. As of 2022, \textit{NYUSIM} has been downloaded more than 100,000 times and is widely used by industry and academia as an alternative to 3GPP SCM. \\
The rest of the paper is organized as follows. Section II provides an overview of the NYUSIM implementation in ns-3. Section III discusses the simulation goals and simulation setup. Section IV discusses the impact of SINR on average end-to-end throughput, latency, and packet drops of wireless modems in all scenarios (UMi, UMa, RMa, and InH), assuming video is transmitted at 50 Mbps using UDP data. Section V describes our numerical results using ns-3 with 3GPP SCM and NYUSIM in the physical (PHY) layer. Section V concludes our work and presents possibilities for future research.

\section{Overview of NYUSIM in ns3}
 This section provides a brief overview of the line of sight (LOS) probability models (which impacts the link SINR experienced by the UE) \cite{Rap2017itap}, large-scale distant dependent path loss, and multipath time cluster spatial lobe (TCSL)-based small-scale fading parameters models used in NYUSIM \cite{samimi20163,samimi2016local} and how they are combined and implemented to generate the overall channel matrix \textbf{H} for NYUSIM in ns-3. The implementation of NYUSIM in ns-3 is similar to the implementation of 3GPP SCM described in \cite{zugno2020implementation}.

\subsection{LOS Probability Models}
LOS probability models are needed to statistically determine the likelihood that an arbitrarily placed mobile device has a line of sight propagation path to the base station or access point. Several different LOS probability models have been reported in the literature for different scenarios \cite{Rap2017itap,sun2015path,jarvelainen2016evaluation}. The NYUSIM LOS probability model is similar to 3GPP SCM, but with the entire formula (i.e., the second equation in Table 7.4.2-1 in \cite{3GPPTR}) being squared to more accurately model tail behavior. The NYUSIM LOS probability model was created based on measurements conducted at 28 and 73 GHz in outdoor environments, and ray tracing was used to determine if the path between the Tx and Rx is in LOS/NLOS \cite{7070688}. Prior work \cite{rappaport2017investigation} shows that 3GPP SCM  has a higher chance of predicting LOS at larger distances (several hundred meters) which is unlikely to be true in the UMi and UMa scenarios based on measurements conducted in New York City \cite{7070688,rappaport2017investigation}. Thus, the UMi and UMa LOS probability models implemented in ns-3 for NYUSIM use the NYU (squared) model based on Tables I and II in \cite{Rap2017itap} as given in \eqref{LosProbUmi} and \eqref{LosProbUma}.
\begin{align}
    \label{LosProbUmi}
    \mathrm{P_{LOS}}(d_{2D}) =  (\min(d_{1}/d_{2D},1)&(1-\exp(-d_{2D}/d_{2})) \nonumber\\&+ \exp(-d_{2D}/d_{2}))^2
\end{align}
\begin{align}
    \label{LosProbUma}
    \mathrm{P_{LOS}}(d_{2D}) =  &((\min(d_{1}/d_{2D},1)(1-\exp(-d_{2D}/d_{2})) \nonumber \\& + \exp(-d_{2D}/d_{2}))(1 + C(d_{2D},h_{UE})))^2
\end{align}
In \eqref{LosProbUmi} $d_{1}=$ 22m, $d_{2}$ = 100m, in \eqref{LosProbUma} $d_{1}=$ 20m, $d_{2}$ = 160m, $C$ is defined in Table 7.4.2 in \cite{3GPPTR} and $h_{UE}$ is the height of the UE in meters. For both \eqref{LosProbUmi} and \eqref{LosProbUma} the values of $d_{1}$, $d_{2}$ are based on measurement data\cite{7070688} and $d_{2D}$ is 2D separation distance between the Tx-Rx in meters.  However, LOS probability models for InH and RMa scenarios do not exist in NYUSIM. Thus, for the InH LOS probability models, we use the 5GCM model from Table III \cite{Rap2017itap} because NYU squared model for UMi and UMa scenarios is similar to that of 5GCM LOS probability models. Whereas, for the RMa scenario, we use the LOS probability models given by 3GPP SCM  in Table 7.4.2 of \cite{3GPPTR}.
\subsection{Large-scale Path Loss Model Parameters}
\begin{figure}
  \begin{subfigure}[b]{0.24\textwidth}
    \includegraphics[width=\textwidth]{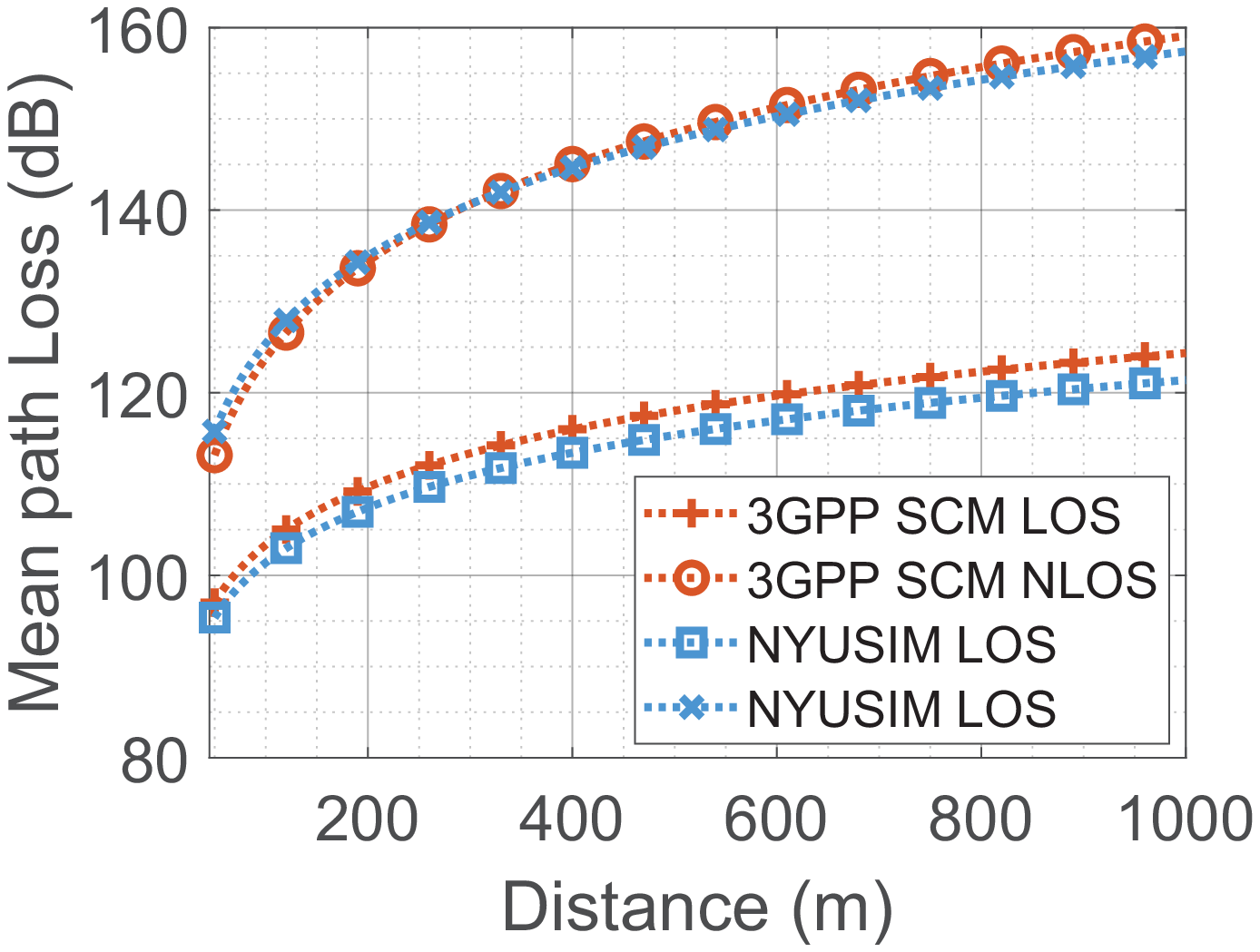}
    \caption{Mean path loss in UMi.}
    \label{fig:subfig1}
  \end{subfigure}
  \begin{subfigure}[b]{0.24\textwidth}
    \includegraphics[width=\textwidth]{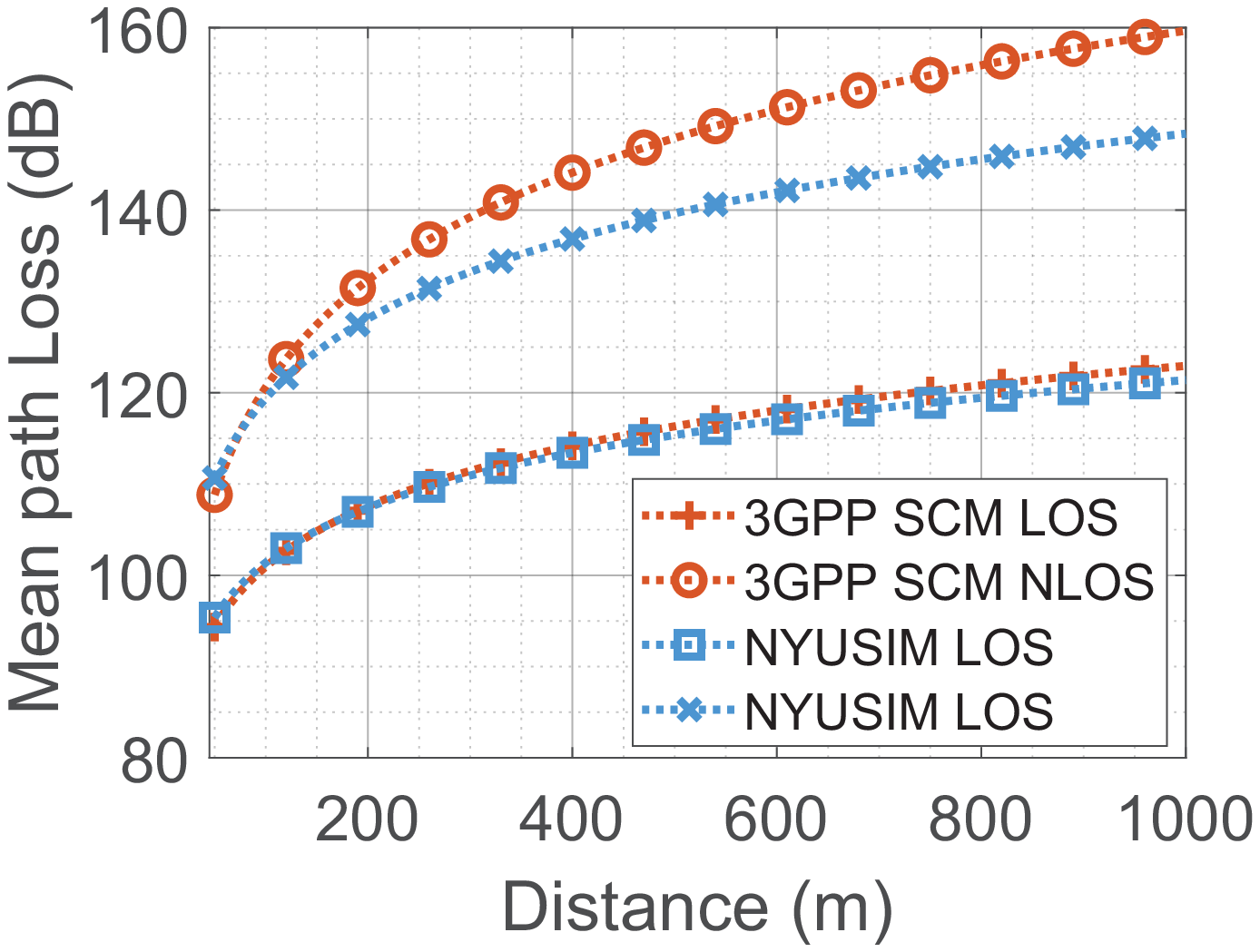}
    \caption{Mean path loss in UMa.}
    \label{fig:subfig2}
  \end{subfigure}
  \vskip\baselineskip
  \begin{subfigure}[b]{0.24\textwidth}
    \includegraphics[width=\textwidth]{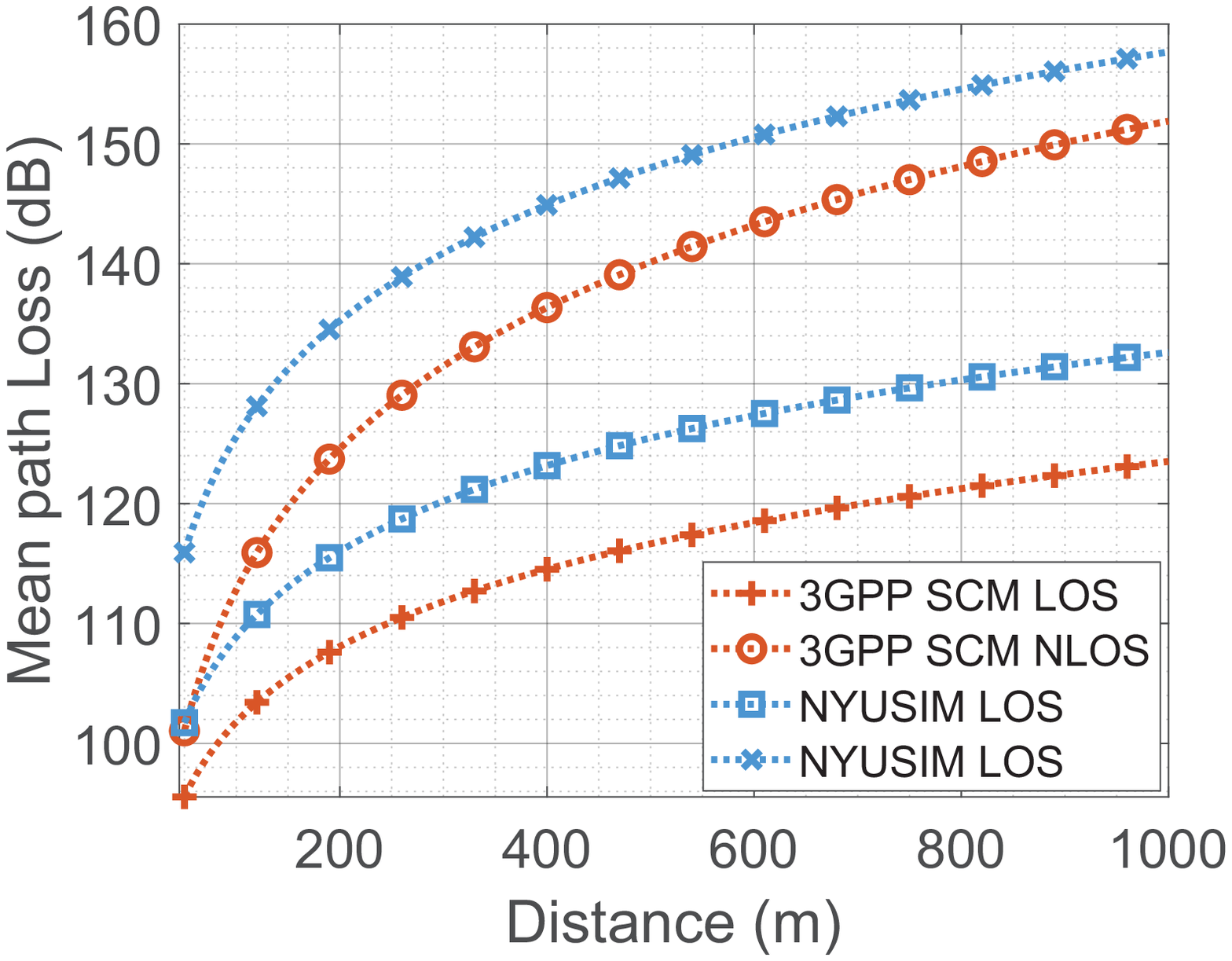}
    \caption{Mean path loss in RMa.}
    \label{fig:subfig3}
  \end{subfigure}
  \begin{subfigure}[b]{0.24\textwidth}
    \includegraphics[width=\textwidth]{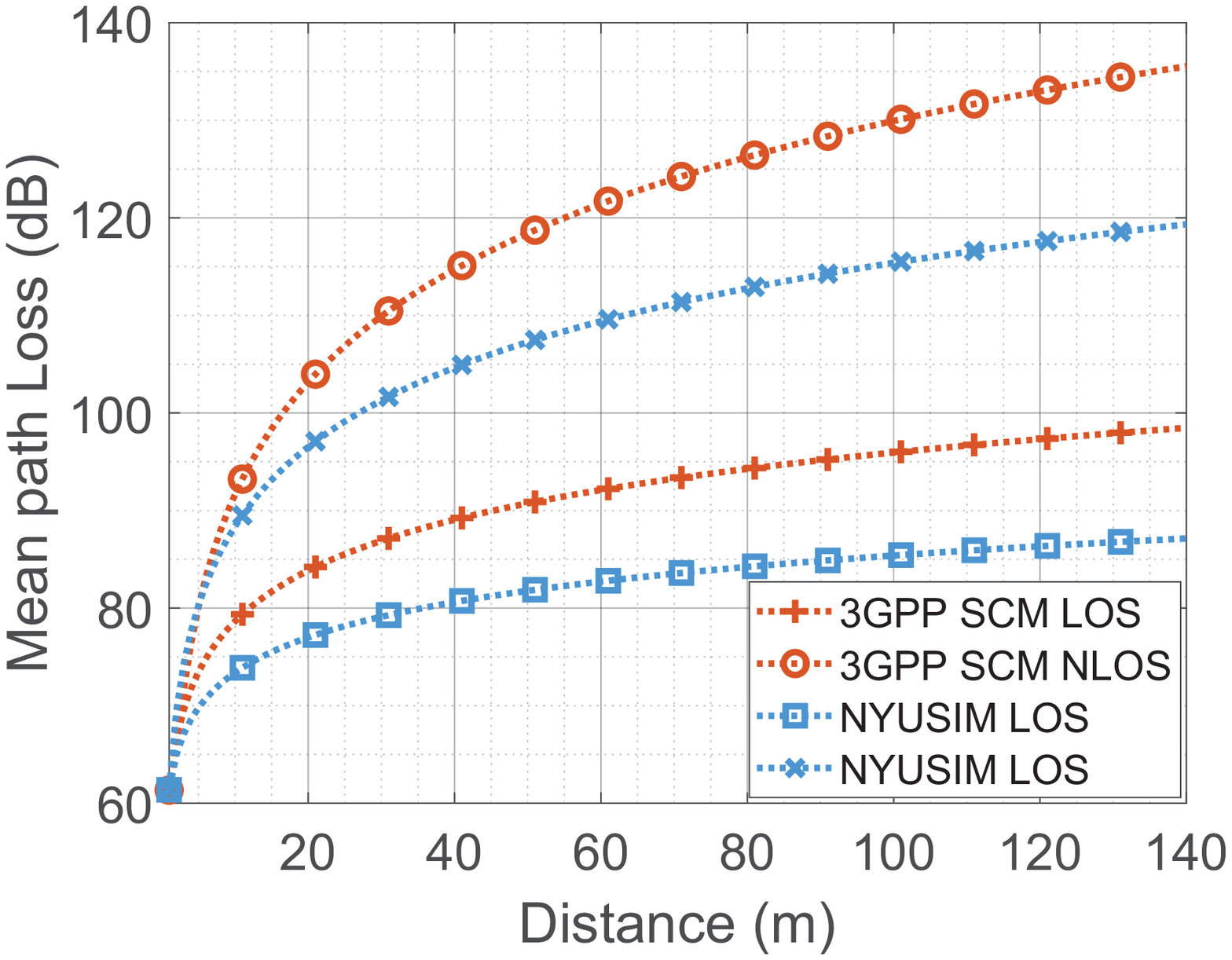}
    \caption{Mean path loss in InH.}
    \label{fig:subfig4}
  \end{subfigure}
  \caption{Mean path loss (dB) vs. distance (m) at 28 GHz computed using 7.4.1-1 in \cite{3GPPTR} for 3GPP SCM and \eqref{ciPathLossModel} for NYUSIM in UMa, UMi, RMa and InH scenarios in LOS and NLOS channel condition.}
  \label{fig:PathLoss}
  \vspace{-0.3in}
\end{figure}
All wireless system models require a large-scale path loss model that accurately predicts received power at a wireless modem as a function of distance from the transmitter. To compute the path loss, regardless of the scenario, NYUSIM uses the close-in free space reference distance (CI) path loss model with a 1 m free space reference distance, which was found to have superior accuracy and less sensitivity to error from measurement campaigns in many different scenarios \cite{sun2016investigation}. NYUSIM uses the CI path loss model given in \eqref{ciPathLossModel}.
\begin{align}
    \label{ciPathLossModel}
    \mathrm{PL^{CI}}&(f,d) [\mathrm{dB}] =  \mathrm{FSPL}(f,\mathrm{1\,m}) [\mathrm{dB}] +  10 \eta \log_{10}(d) \nonumber \\&+ \mathrm{AT} [\mathrm{dB}] + \mathrm{O2I} [\mathrm{dB}] + \mathrm{FL} [\mathrm{dB}] + \chi_{\sigma}^{CI} [\mathrm{dB}]
\end{align}
where $f$ is the frequency in GHz, $d$ denotes the 2D T-R separation distance in meters, and $d \geq 1$ m, $\mathrm{FSPL}(f,1 \; \mathrm{m})$ represents the free space path loss at a T-R separation of 1 m at carrier frequency $f$, $\eta$ denotes the path loss exponent (PLE). AT represents the atmospheric attenuation, and O2I indicates the outdoor to indoor loss caused due to the propagation of the signal from the outdoor to the indoor environment \cite{9838919}. O2I loss is implemented using the high and low loss parabolic models for building penetration loss  \cite{haneda20165g},\cite{5GCMChannelModel}, and FL represents the foliage loss. $\chi_{\sigma}^{CI}$ is a zero-mean Gaussian random variable with $\sigma$ as the standard deviation in dB to represent shadow fading about the distant-dependent mean path loss value. The expression for FL is based on measurements conducted at 73 GHz in NYU MetroTech Commons Courtyard in Brooklyn \cite{7247347} and is given as follows.
\begin{align}
    \label{foliageLoss}
    \mathrm{FL} [\mathrm{dB}] =  \alpha[\mathrm{dB/m}] \times d[\mathrm{m}]
\end{align}
where $\alpha$ is the attenuation factor in dB/m and can range from 0-10 dB \cite{7247347}.
The PLE and $\chi_{\sigma}^{CI}$ values given in \eqref{ciPathLossModel} for LOS and NLOS in the four scenarios shown in Fig. \ref{fig:PathLoss} were empirically determined by curve fitting using measurement data at 28 GHz, 38 GHz, 73 GHz, and 140 GHz \cite{rappaport2013millimeter,maccartney2015indoor,maccartney2016millimeter,maccartney2017rural,ju20203,9558848}. Since NYUSIM is based on measurements just at a few different frequency bands, it becomes necessary to interpolate the large-scale path loss model parameters such as PLE and $\chi_{\sigma}^{CI}$ for any arbitrary frequency of interest. We neglect oxygen absorption (which could easily be accounted for if desired by using the data in \cite{9450810}) and do a linear interpolation between the PLE and $\chi_{\sigma}^{CI}$ values obtained at, for example, 28 GHz and 140 GHz, as given in \eqref{linearInterpolation}.
\begin{align}
    \label{linearInterpolation}
    \resizebox{.9\hsize}{!}{$
    	p(f)=\left\{\begin{matrix}
		p(28) & ,f \leqslant 28\\ 
		\frac{p(140)-p(28)}{140-28}f+\frac{5p(28)-p(140)}{4} & , 28 < f < 140\\ 
		p(140) & , f \geqslant 140
	\end{matrix}\right.
	$}
\end{align}
where $p(f)$ denotes the PLE or $\chi_{\sigma}^{CI}$ value at frequency $f$ in GHz, $p(28)$ and $p(140)$ represent the PLE or $\chi_{\sigma}^{CI}$ at 28 GHz and 140 GHz, respectively.
Fig. \ref{fig:PathLoss} shows the mean path loss values computed in all scenarios at 28 GHz for 3GPP SCM using
7.4.1-1 in \cite{3GPPTR} and NYUSIM using \eqref{ciPathLossModel} for LOS and NLOS channel condition.
The height of the UE is fixed to 1.6 m in UMi, UMa, RMa and InH scenarios. The gNB height is set to 10 m and 3 m for UMi and InH scenarios. For both the UMa and RMa scenario, the gNB height is 25 m.

\subsection{Small-scale Multipath Models}
Multipath (MP) models capture the time-varying nature of the wireless channel and help in statistically reproducing the channel impulse responses (CIR). The CIR comprises the angles of arrival, time of arrival (over tens or hundreds of nanoseconds), and the amplitude of the multipath components (MPCs). The 3GPP SCM use a joint delay angle probability density function \cite{3GPPTR}, and NYUSIM uses a time cluster (TC) and spatial lobes (SL) (TCSL) approach to characterize the temporal and spatial properties \cite{samimi20163} of the MPCs. The TCSL approach is borne out of extensive measurement campaigns by NYU WIRELESS from 2011-2022. TC comprises MPCs traveling closely in time (tens or hundreds of nanoseconds) \cite{samimi20163}, and SL denotes the direction of arrival/departure of most energy over the azimuth or elevation plane \cite{samimi20163}.
\subsection{Channel Matrix H for NYUSIM}
A channel matrix \textbf{H} consists of all paths between the transmit antennas at the transmitter and receive antennas at the receiver. We implement a 3GPP-like
channel matrix for NYUSIM in ns-3 based on 3GPP TR 38.901 \cite{rebato2018multi} and it is similar to the 3GPP channel matrix \cite{3GPPTR} and is expressed as follows.
\vspace{-0.04in}
\begin{align}
    \label{nyuChannelMatrix}
    \boldsymbol{H(\tau,t)} = \sum_{i = 1}^{M} H_{u,s,m}(t) \delta(\tau - \tau_{i})
\end{align}
where $M$ denotes the total number of MPCs, $i$ represents the $i^{th}$ MP, and $\tau_{i}$ is the delay of the $i^{th}$ MP. The channel coefficients $H_{u,s,m}(t)$ at time $t$ are computed as follows.
\begin{equation}
\begin{aligned}
    \label{nyuChannelCoefficient}
    &H_{u,s,m}(t) =  \alpha_{m}
    \begin{bmatrix}
    F_{rx,u,\theta}(\theta_{m,ZOA},\phi_{m,AOA}) \\
    F_{rx,u,\phi}(\theta_{m,ZOA},\phi_{m,AOA})
    \end{bmatrix}^\top \\ \times 
    &\begin{bmatrix}
    exp(j\phi_{m}^{\theta\theta}) & \sqrt{\frac{1}{K_{m_{\theta,\phi}}}} \exp(j \phi_{m}^{\theta\phi})\\
     \sqrt{\frac{1}{K_{m_{\phi,\theta}}}}exp(j \phi_{m}^{\phi\theta}) & \sqrt{\frac{1}{K_{m_{\phi,\phi}}}} \exp(j \phi_{m}^{\phi\phi})
     \end{bmatrix}\\ \times
    &\begin{bmatrix}
     F_{tx,s,\theta}(\theta_{m,ZOD},\phi_{m,AOD}) \\
     F_{tx,s,\phi}(\theta_{m,ZOD},\phi_{m,AOD})
     \end{bmatrix}  \\\times
     & \exp\left[\frac{j2\pi(\hat{r}^T_{rx,m}. \Bar{d}_{rx,u})}{\lambda_{o}}\right]
     \exp\left[\frac{j2\pi(\hat{r}^T_{tx,m}. \Bar{d}_{tx,s})}{\lambda_{o}}\right]\\
    %
\end{aligned}
\end{equation}\\
where $m$ denotes the $m^{th}$ MP and $\alpha_{m}$ is the received amplitude of the $m^{th}$ MP in linear scale. $\alpha_{m}$ is generated using the TCSL approach. Cross-polarization ratios (XPD), $K_{m_{\theta,\phi}}, K_{m_{\phi,\theta}}, K_{m_{\phi,\phi}}$ are in linear scale are computed using NYUSIM \cite{nyusim}. Furthermore, {$\phi_{m}^{\theta\theta}, \phi_{m}^{\theta\phi}, \phi_{m}^{\phi\theta}, \phi_{m}^{\phi\phi}$} are generated for each MP $m$ in the interval $(0,2\pi)$. The description of other parameters in \eqref{nyuChannelCoefficient} is given in \cite{3GPPTR}.

\section{Simulation Goal and Setup}
We aim to study the impact of SINR on the end-to-end throughput, latency, and packet drop observed by a wireless modem operating at 28 GHz with a bandwidth of 100 MHz in UMi, UMa, RMa, and InH scenarios in LOS and NLOS channel conditions using 3GPP SCM and NYUSIM. Video is transmitted from a remote server using UDP at 50 Mbps (4K/8K video, low-end AR/VR applications) over an NR network (ns-3 mmWave module). Simulations performed in this work consider one gNB and one UE. The UE is fixed at 100 m from the gNB. The gNB transmits at a power of 30 dBm. The simulations are run by fixing the link to a deterministic LOS/NLOS state. We run $50$ realizations for each combination of channel model (NYUSIM/3GPP SCM) and channel condition (fixed LOS/NLOS) in each scenario. In each realization (duration of 9 seconds), HARQ is enabled, and blockage is disabled for both 3GPP SCM and NYUSIM.

\section{Impact of SINR on Average end-to-end throughput, latency, and packet drops}
The mmWave module in ns-3 abstracts most PHY layer operations at the symbol level \cite{lagen2020new}. Furthermore, for a given target error probability and  SINR, the Adaptive Modulation and Coding (AMC) algorithms select the highest possible Modulation and Coding Scheme (MCS) \cite{lagen2020new}. The physical layer throughput at the receiver is expressed as follows:
\begin{align}
    \label{PhyLayerThroughput}
    S =  L/T_{slot}
\end{align}
where $S$ indicates the throughput at the physical layer in Mbps, $L$ is the size of the Transport Block (TB) in bits, and $T_{slot}$ denotes the slot duration in microseconds.
When the SINR is high, a higher MCS (i.e., exhibiting a higher modulation order and/or code rate) is selected. In turn, this leads to a larger TB size and thus, $S$ being greater than the application source rate ($R$). 
In this case, there is generally no buffering at the Radio Link Control layer (RLC), thus, no packets are dropped, and a higher average end-to-end throughput is achieved (\textit{Throughput is inversely proportional to packet drops}). When the SINR is low, a smaller MCS is picked, thus leading to a smaller TB size; this causes a lower $S$. When the SINR is particularly low ($<$ 0 dB), irrespective of the selected MCS, there are packet errors that lead to Hybrid Automatic Repeat reQuest (HARQ) retransmissions. In addition to packet errors, packet drops happen due to buffering at the RLC layer due to $S$ being smaller than $R$. Using HARQ, one can reduce the average end-to-end packet drop and increase the average end-to-end throughput at the cost of increased average end-to-end latency. The increase in the latency is due to re-transmissions or the use of various combining techniques (such as incremental redundancy). 
Furthermore, when $S > R$, the \textit{application layer throughput} saturates after a specific SINR value. This implies that increasing the SINR beyond a certain value does not increase the application layer throughput any further \cite{shakkottai2003cross}. This can happen for two reasons: either the target $R$ is already achieved, or the SINR is high enough for the highest MCS index to be used.
For instance, let us consider a sample scenario simulated in mmWave module on ns-3 using NYUSIM where $R$ is 50 Mbps, the sub-carrier spacing (SCS) is 60 kHz (slot duration 250 us), the observed SINR is 1.3 dB, and the selected MCS is 0 with a TB size of 500 bytes. Using \eqref{PhyLayerThroughput}, $S$ is computed as 16 Mbps. Thus, one can observe that $S < R$. In this scenario, there is a significant average end-to-end packet drop due to the buffering at RLC. This reduces the average end-to-end throughput. However, techniques such as HARQ reduce the average end-to-end packet drop and increase the average end-to-end throughput at the cost of increased latency. Let us consider two more sample scenarios simulated in the mmWave module on ns-3 using NYUSIM, with the same $R$ and SCS configuration. The observed SINRs are 5.6 dB and 58 dB, and the corresponding TB size in bytes is 1600 and 12761. Using \eqref{PhyLayerThroughput}, the computed $S$ is 51.2 Mbps and 408.3 Mbps, respectively. Thus in both the cases, $S > R$. This implies that for a given application rate, the application layer throughput saturates after a particular SINR value for a given $R$ \cite{shakkottai2003cross}. 

\section{Results}
\begin{figure}
  \begin{subfigure}[b]{0.24\textwidth}
    \includegraphics[width=\textwidth]{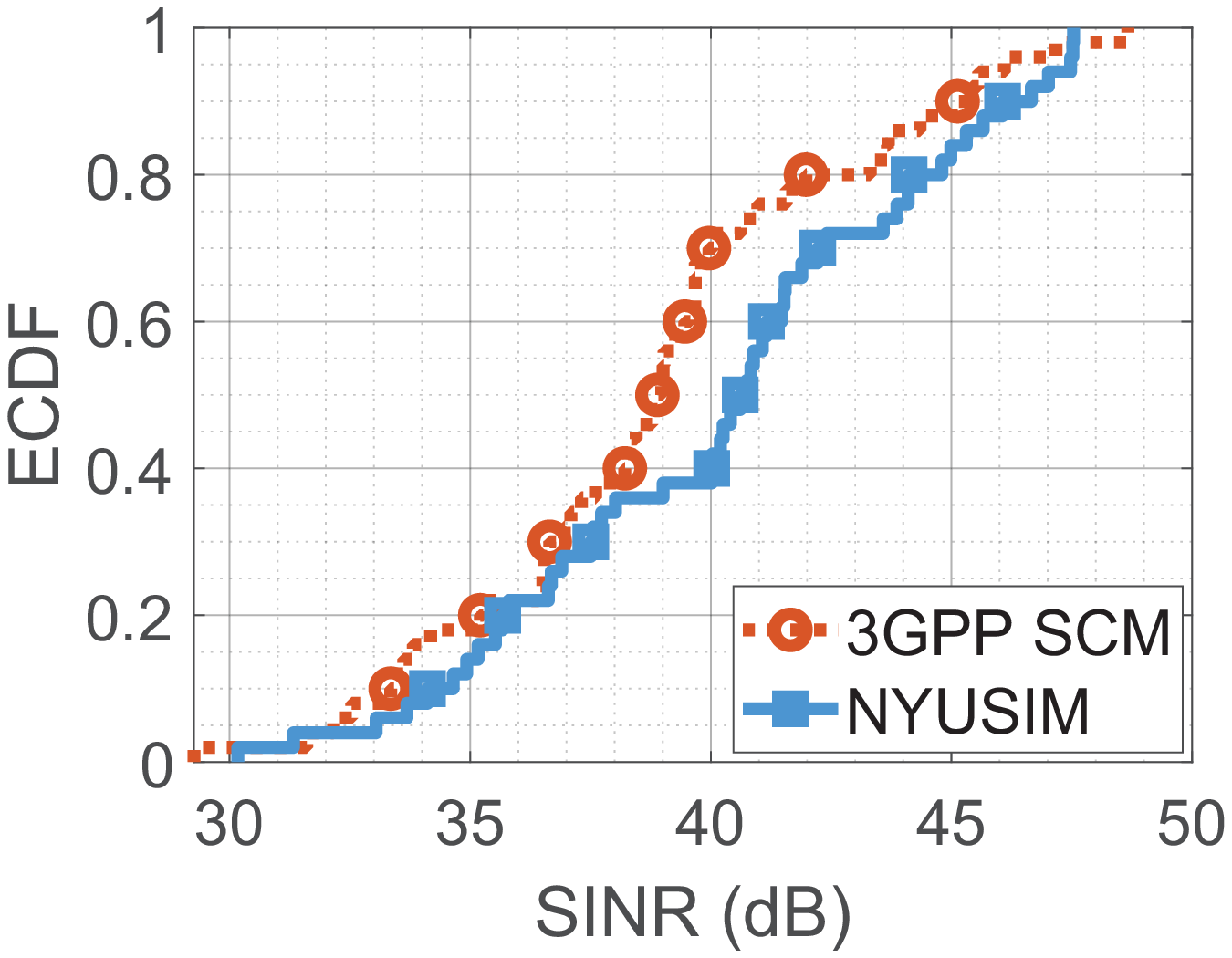}
    \caption{ECDF of SINR for UMi in LOS.}
    \label{fig:subfig1}
  \end{subfigure}
  \begin{subfigure}[b]{0.24\textwidth}
    \includegraphics[width=\textwidth]{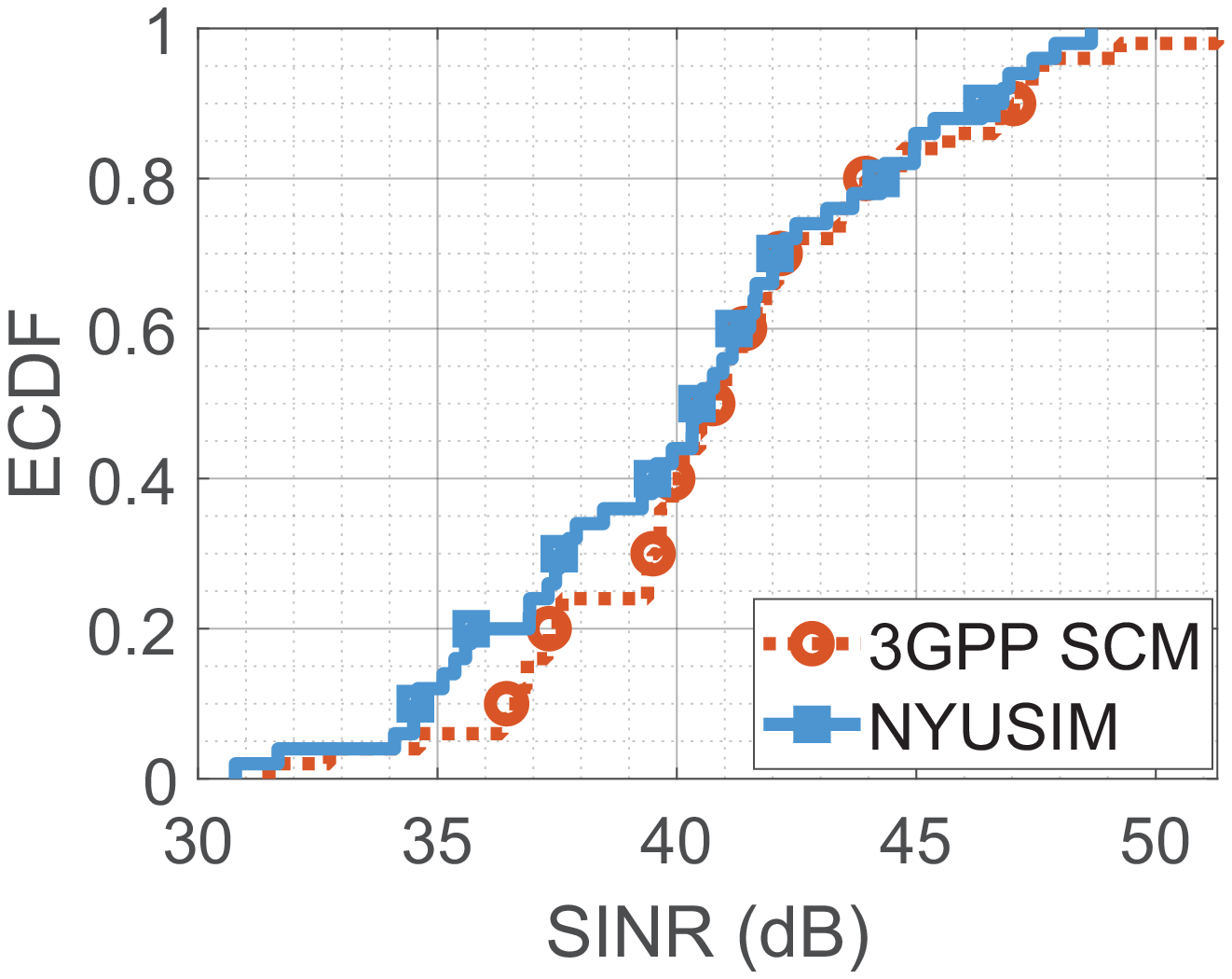}
    \caption{ECDF of SINR for UMa in LOS.}
    \label{fig:subfig2}
  \end{subfigure}
  \vskip\baselineskip
  \begin{subfigure}[b]{0.24\textwidth}
    \includegraphics[width=\textwidth]{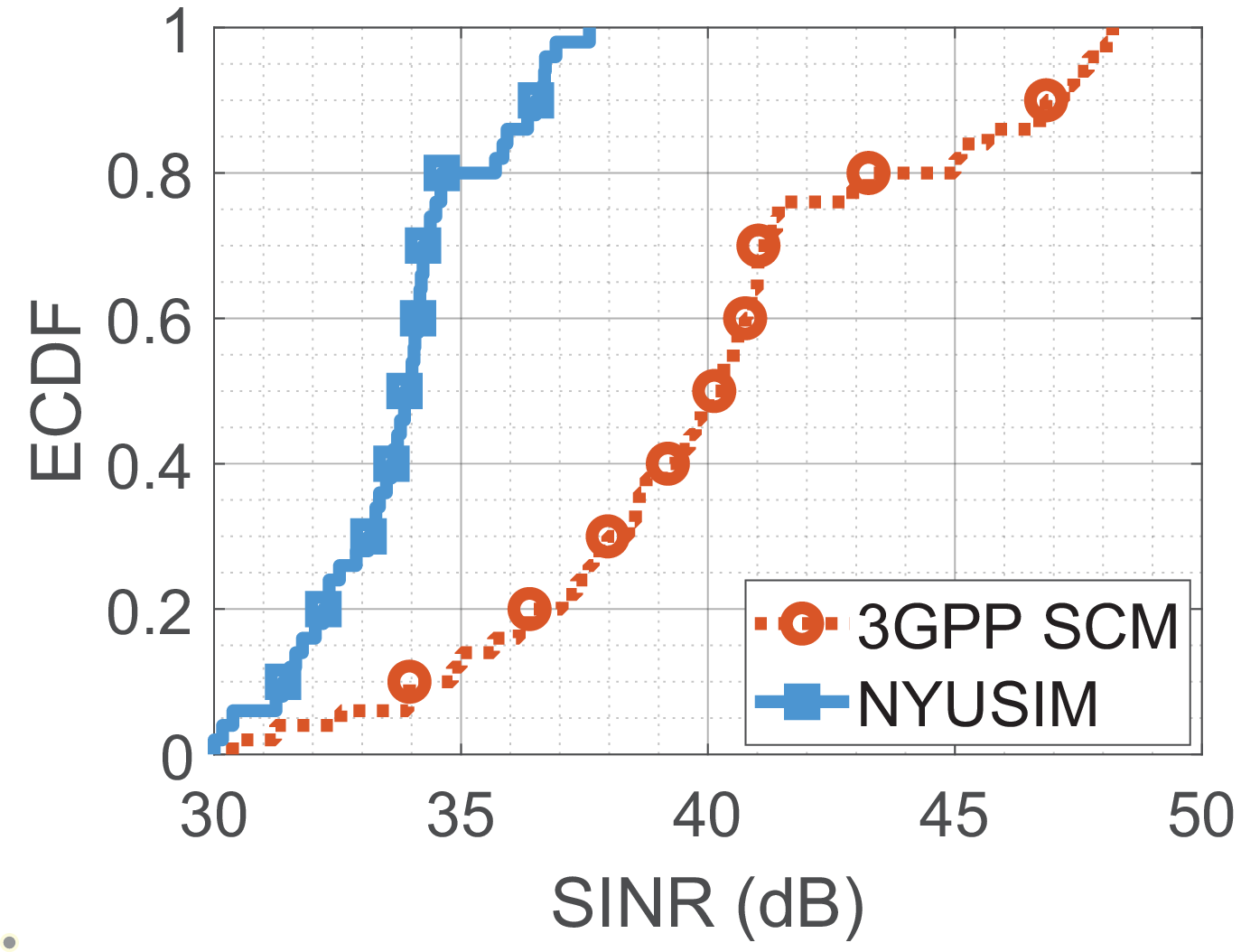}
    \caption{ECDF of SINR for RMa in LOS.}
    \label{fig:subfig3}
  \end{subfigure}
  \begin{subfigure}[b]{0.24\textwidth}
    \includegraphics[width=\textwidth]{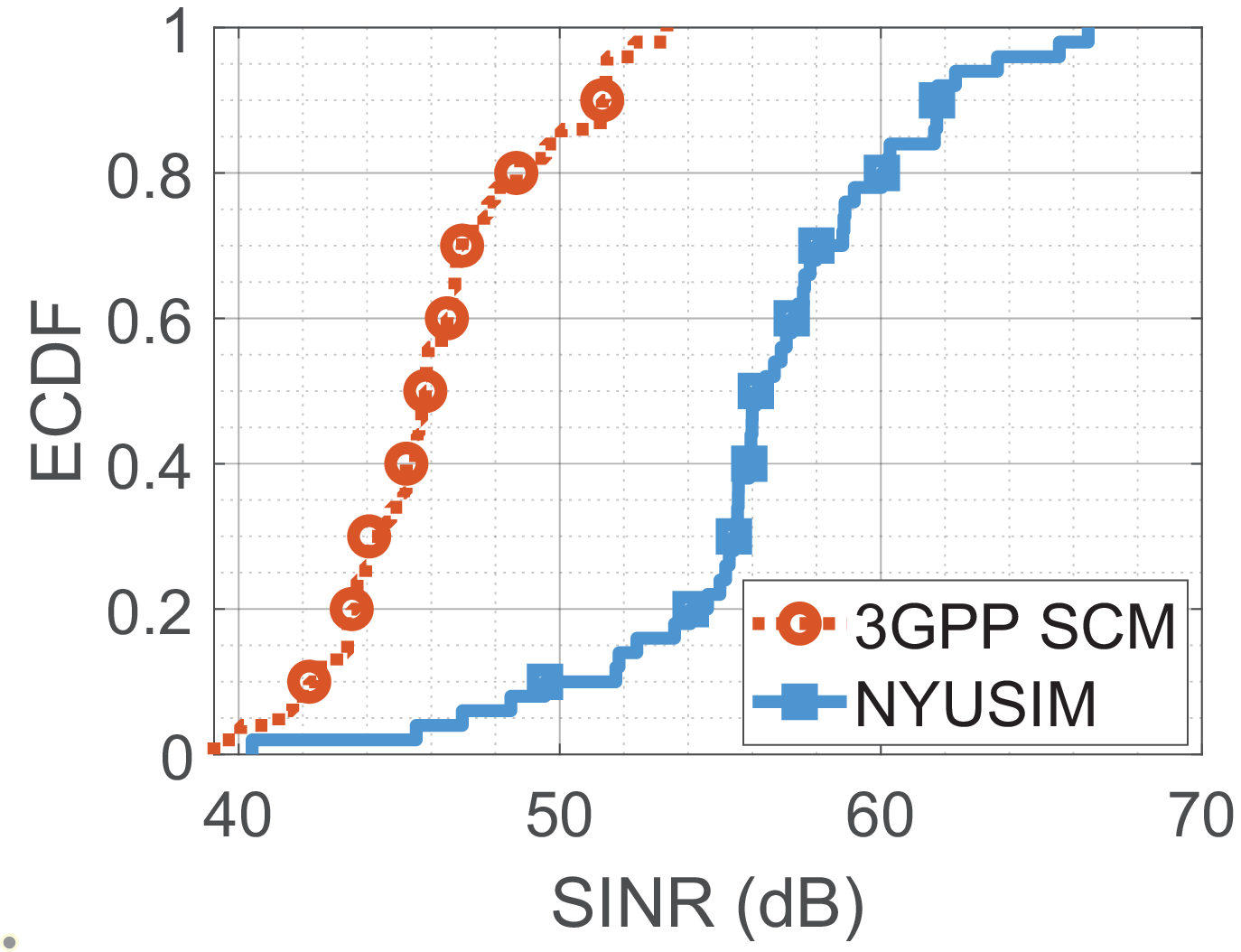}
    \caption{ECDF of SINR for InH in LOS.}
    \label{fig:subfig4}
  \end{subfigure}
  \caption{Empirical cumulative distribution function (ECDF) of SINR in LOS condition for 3GPP SCM vs. NYUSIM in UMi, UMa, RMa, and InH for 28 GHz and 100 MHz bandwidth.}
  \label{fig:LosCDFSinr}
  \vspace{-0.25in}
\end{figure}

\begin{figure}
  \begin{subfigure}[b]{0.24\textwidth}
    \includegraphics[width=\textwidth]{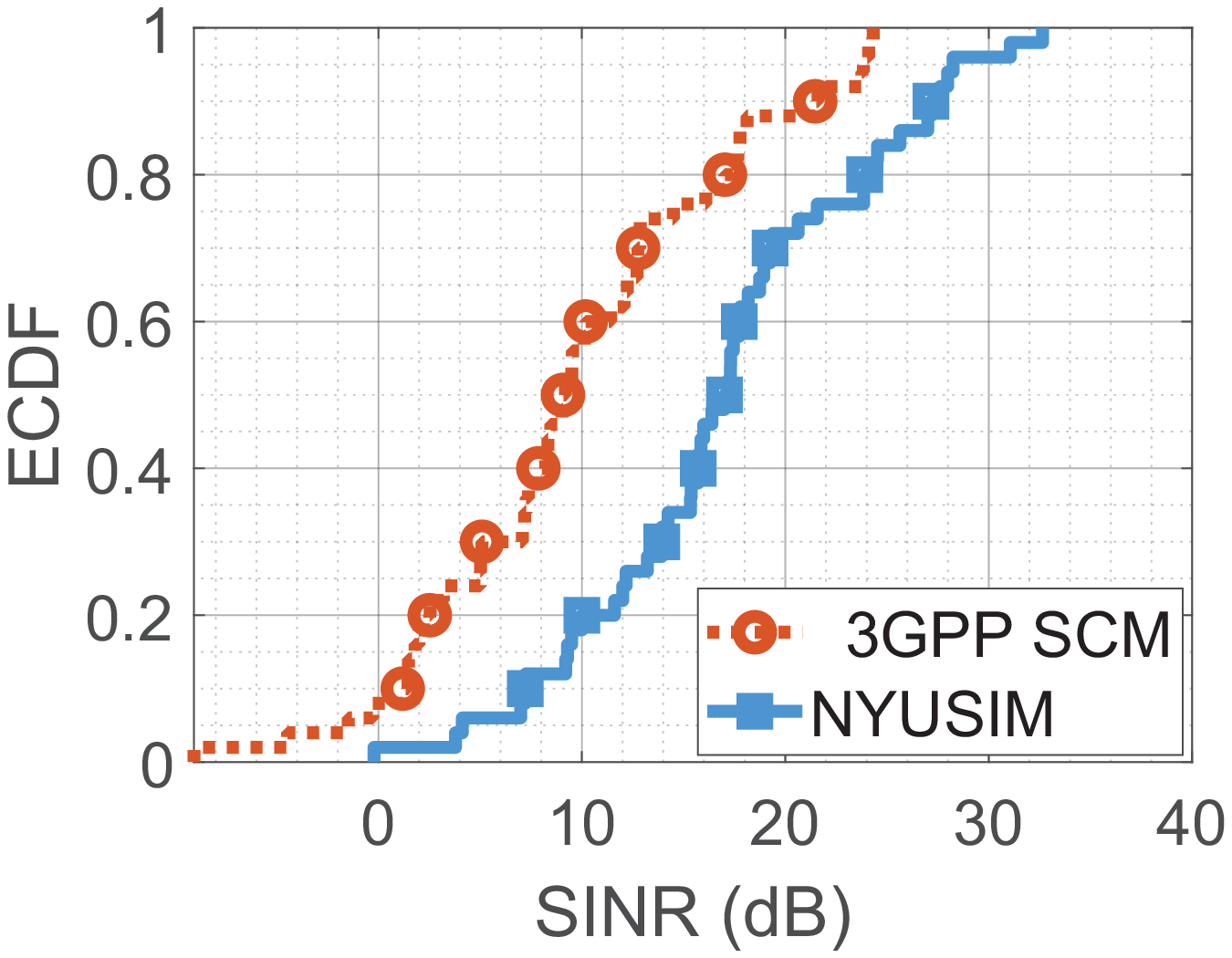}
    \caption{ECDF of SINR for UMi in NLOS.}
    \label{fig:subfig1}
  \end{subfigure}
  \begin{subfigure}[b]{0.24\textwidth}
    \includegraphics[width=\textwidth]{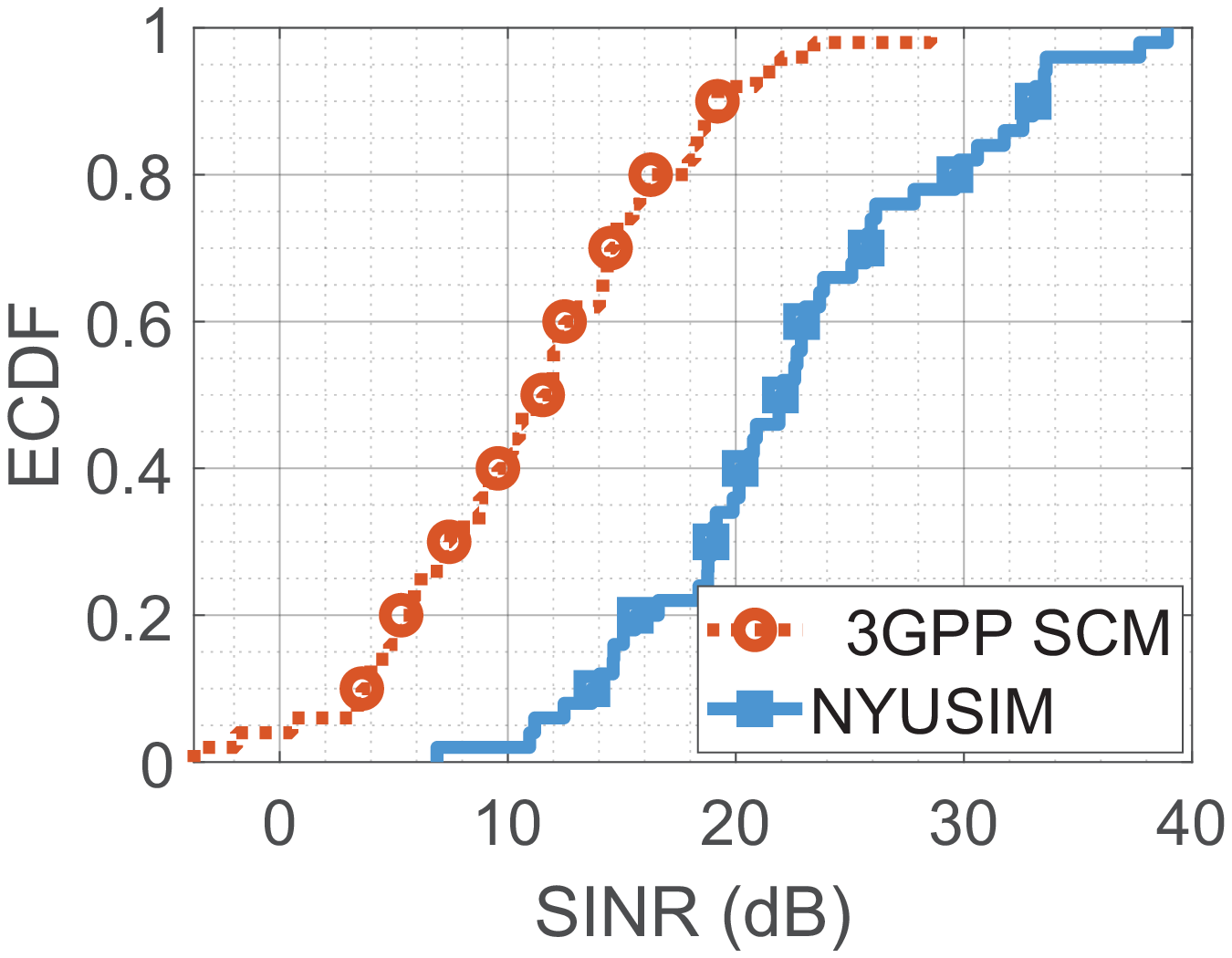}
    \caption{ECDF of SINR for UMa in NLOS.}
    \label{fig:subfig2}
  \end{subfigure}
  \vskip\baselineskip
  \begin{subfigure}[b]{0.24\textwidth}
    \includegraphics[width=\textwidth]{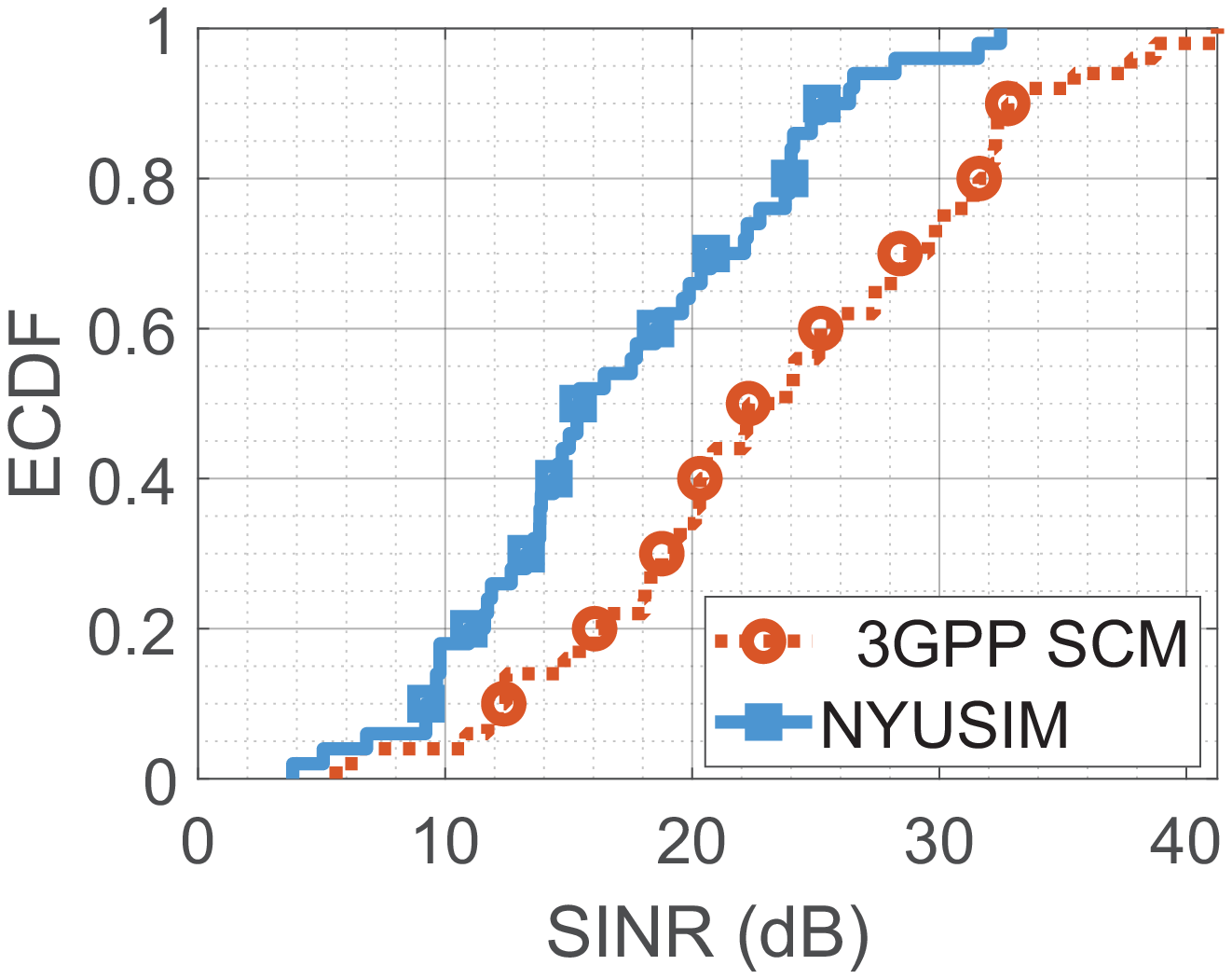}
    \caption{ECDF of SINR for RMa in NLOS.}
    \label{fig:subfig3}
  \end{subfigure}
  \begin{subfigure}[b]{0.24\textwidth}
    \includegraphics[width=\textwidth]{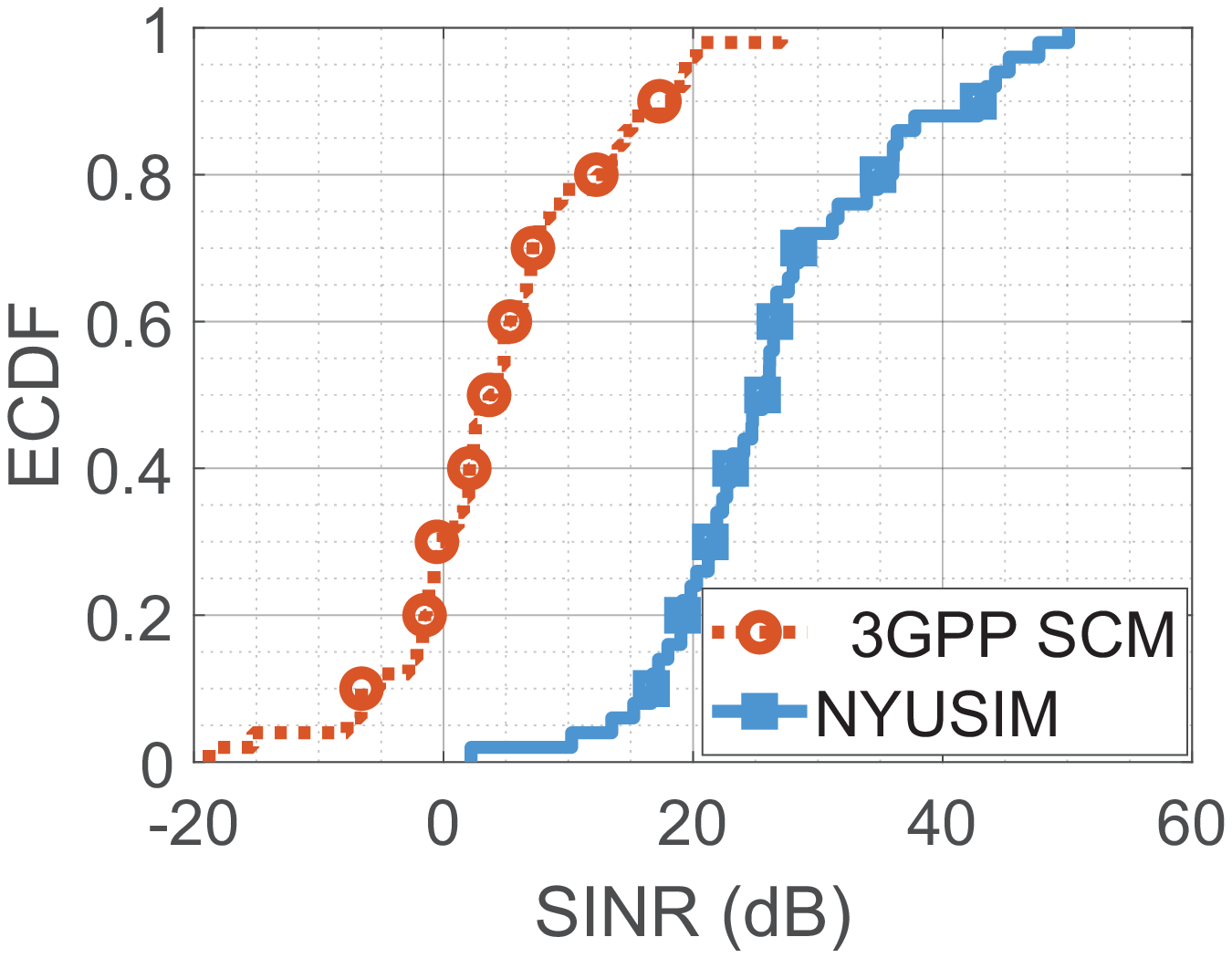}
    \caption{ECDF of SINR for InH in NLOS.}
    \label{fig:subfig4}
  \end{subfigure}
  \caption{Empirical cumulative distribution function (ECDF) of SINR in NLOS condition for 3GPP SCM vs. NYUSIM in UMi, UMa, RMa, and InH for 28
GHz and 100 MHz bandwidth.}
  \label{fig:NlosCDFSinr}
  \vspace{-0.25in}
\end{figure}


We evaluate and compare the performance of the wireless modem in UMi, UMa, RMa, and InH for LOS and NLOS using 3GPP SCM and NYUSIM (the simulation parameters are described in section III) in terms of the following metrics:
\begin{itemize}
    \item \textit{SINR}: It is the ratio of the received signal power and the sum of noise and interference.
    \item \textit{Throughput}: It is the total number of bits received by the UE per second.
    \item \textit{Latency}: It is the time difference between the generation of a packet at the UDP layer at gNB and the reception of the packet at the UDP layer of the UE. This includes both transmission and queuing times. 
    \item \textit{Packet drops}: It is the difference between the number of packets transmitted at the UDP layer by the gNB and the number of packets received at the UDP layer by the UE.
\end{itemize}

\renewcommand{\arraystretch}{1}
\begin{table*}[t]
  \centering
  \caption{\label{results}\centering Comparison of average (Avg.) end-to-end throughput, latency, and packet drop in UMi, UMa, RMa, and InH scenarios for LOS and NLOS at the UDP Layer of the UE assuming downlink video transmission at 50 Mbps via UDP (packet size 62500 bytes) using 3GPP SCM and NYUSIM.}
    \begin{tabular}{|c|c|c|c|c|c|c|c|c|c|c|c|c|}
    \hline
    \multirow{3}{*}{\centering Scenario} & \multicolumn{4}{c|}{Avg. end-to-end throughput (Mbps)} & \multicolumn{4}{c|}{Avg. end-to-end latency (ms)} & \multicolumn{4}{c|}{Avg. end-to-end packet drop (\%)} \\
\cline{2-13}          & \multicolumn{2}{c|}{LOS} & \multicolumn{2}{c|}{NLOS} & \multicolumn{2}{c|}{LOS} & \multicolumn{2}{c|}{NLOS} & \multicolumn{2}{c|}{LOS} & \multicolumn{2}{c|}{NLOS} \\
\cline{2-13}          & 3GPP  & NYU   & 3GPP  & NYU   & 3GPP  & NYU   & 3GPP  & NYU   & 3GPP  & NYU   & 3GPP  & NYU \\
    \hline
    UMi   & 48.1  & 48.1  & 40.2  & 47.1  & 4.1   & 4.1   & 811.4 & 123.7  & 0     & 0     & 16.4  & 2 \\
    \hline
    UMa   & 48.1  & 48.1  & 43.5  & 48.1  & 4.1   & 4.1   & 588.7 & 4.7   & 0     & 0     & 9.5   & 0 \\
    \hline
    RMa   & 48.1  & 48.1  & 48.1  & 48.1  & 4.1   & 4.1   & 19.2     & 19.3   & 0     & 0     & 0   & 0 \\
    \hline
    InH   & 48.1  & 48.1  & 27.5  & 47.7  & 4.1   & 4.1   & 2071  & 49.2  & 0     & 0     & 42.9  & 1 \\
    \hline
    \end{tabular}%
  \label{tab:Results}%
  \vspace{-0.1in}
\end{table*}%
\renewcommand{\arraystretch}{1}

\subsubsection{\textbf{SINR}} 
Fig. \ref{fig:LosCDFSinr} and Fig. \ref{fig:NlosCDFSinr} present the plots of the ECDF of the SINR for LOS and NLOS channel conditions in all the scenarios considered in this work. As mentioned earlier, we have considered only one gNB and one UE. Therefore, the SINR in the simulations depends only on the path loss (including $\chi_{\sigma}^{CI}$). From Fig. \ref{fig:LosCDFSinr}, we can observe that in the LOS channel condition, SINR $\geq$ 30 dB for both 3GPP SCM and NYUSIM in all the scenarios due to lower path loss in all scenarios. A lower path loss coupled with beamforming in the direction of the LOS path leads to a stronger received power and, thus, a higher SINR. On the other hand, Fig. \ref{fig:NlosCDFSinr} indicates that under NLOS channel conditions, SINR $\geq$ 0 dB for NYUSIM in all scenarios, except in one or two simulation runs the SINR $\leq$ 0 dB for UMi scenario. In contrast, for the 3GPP SCM, the SINR is $<$ 0 dB for UMa, UMi and InH scenarios and $>$ 0 dB for RMa scenarios. In both the channel models (3GPP SCM and NYUSIM), the reason for SINR being $\leq$ 0 dB is that beamforming gain in the direction of strong multipath cannot compensate for the path loss leading to lower received signal strength at the UE. For instance, in 3GPP SCM, for the InH scenario, the computed path loss is very high compared to the path loss for the UMi scenario (Fig. \ref{fig:PathLoss}), and beamforming gain is unable to compensate for the higher path loss which causes a degradation in SINR.  

\subsubsection{\textbf{Average end-to-end throughput}}
From Table. \ref{results}, we can see that the average end-to-end throughput is $\sim$ 48.1 Mbps when we fix the channel condition to LOS for 3GPP SCM and NYUSIM in all the scenarios as the SINR $\geq 30$ dB (Fig. \ref{fig:LosCDFSinr}) leading to $S$ $>$ $R$ (as discussed in section IV). For NLOS, the average end-to-end throughput in NYUSIM is 47.1 Mbps, 48.1 Mbps, 48.1 Mbps, and 47.7 Mbps for UMi, UMa, RMa, and InH scenarios. For UMa and RMa scenarios, the SINR is high enough to support $S$ $>$ $R$. There is a slight drop in throughput for the UMi and InH scenario compared to that of the UMa scenario because in one-two simulation runs, the SINR is low, which causes $S<R$. On the other hand, 3GPP SCM  exhibits an average end-to-end throughput of 40.2 Mbps, 43.5 Mbps, 48.1 Mbps, and 27.5 Mbps in NLOS channel conditions for UMi, UMa, RMa, and InH scenarios. The RMa channel supports the highest throughput because $S$ $>$ $R$. Whereas, for the InH scenario, the throughput is the lowest because, in $\sim$ 40\% of the simulations run SINR $\leq$ 0 dB causing $S$ $<$ $R$.

\subsubsection{\textbf{Average end-to-end latency}} 
Table. \ref{results} shows the average end-to-end latency. 3GPP SCM and NYUSIM result in the same average end-to-end latency of 4.1 ms in LOS condition in all scenarios as SINR $\geq 30$ dB, leading to the selection of the highest MCS and lowest average end-to-end latency. The average end-to-end latency in NLOS for NYUSIM is 123.68 ms, 4.7 ms, 19.3 ms, and 49.23 ms in UMi, UMa, RMa, and InH scenarios, respectively. We achieve a lower latency for the UMa as $S \geq R$. For RMa scenario the latency increases compared to UMa as the RLC buffer becomes full due to $S < R$ in one simulation run. On the other hand, for UMi and InH scenarios, HARQ comes into play in a few simulation runs as $S \leq R$. For 3GPP SCM, the average end-to-end latency is 811.4 ms, 588.7 ms, 19.2 ms, and 2071 ms for UMi, UMa, RMa, and InH scenarios, respectively. In the RMa scenario, $S \geq R$, whereas in the UMi, UMa, and InH scenario, the SINR $< 0 $ dB for few of the simulation runs. In these scenarios, HARQ is triggered, which causes an increase in latency. Furthermore, latency is highest in InH due to SINR $\leq$ 0 dB in more simulation runs compared to UMi and UMa. Thus, the number of retransmissions required increases causing the latency to increase drastically.

\subsubsection{\textbf{Average end-to-end packet drop}} 
From Table. \ref{results}, it is clear that in LOS, 3GPP SCM and NYUSIM exhibit no packet drop (as there is no buffering at the RLC) in all the scenarios due to the high SINR values ($\geq30$ dB for all channel realizations as seen in Fig .\ref{fig:LosCDFSinr}). In NLOS, the packet drop for the 3GPP SCM are 16.4\%, 9.5\%, 0\%, and 42.9\% for UMi, UMa, InH, and RMa scenarios. In InH, in $\sim$ 25\% of the simulation runs, the received SINR values are $\leq-10$ dB leading to the highest number of packet drops due to buffering at the RLC layer. Furthermore, in UMi and UMa scenarios, SINR in 10\% and 5\% of the realizations is $\leq$ 0 dB, which leads to packet drops ($S \leq R$), however, less than that of InH scenario. Similarly, using NYUSIM, the average end-to-end packet drops are 2\%, 0\%, 0\%, and 1\% in UMi, UMa, RMa, and InH scenarios, respectively. In UMa and RMa scenarios, SINR is sufficiently high to avoid buffering at the RLC in most of the simulation runs for all scenarios. In contrast, in UMi and InH, the SINR is low in a few realizations, as stated earlier, causing few packet drops.

\section{Conclusion and Future Work}
This article compares the end-to-end throughput, latency, and packet drop of a wireless modem in UMi, UMa, RMa, and InH using 3GPP SCM and NYUSIM under LOS and NLOS channel conditions. In LOS and NLOS channel conditions for the UMi, UMa, and InH scenarios, NYUSIM exhibits a similar or higher SINR than 3GPP SCM due to lower path loss and greater directionality. A lower path loss coupled with beamforming in the direction of the strong multipaths leads to a stronger received power and, thus, a higher SINR for NYUSIM than 3GPP SCM. 
Simulation results show no packet drop, maximum throughput (48.1 Mbps), and lowest latency (4.1 ms) is achieved in LOS for all scenarios mentioned above using 3GPP SCM and NYUSIM. The RMa scenario in 3GPP SCM  and NYUSIM for NLOS exhibits similar performance because the physical layer throughput exceeds the application source rate due to a larger SINR. The InH scenario in 3GPP SCM  for NLOS shows the lowest throughput (27.5 Mbps), highest latency (2071 ms), and packet drops (42.9\%) due to extremely low SINR. Thus, to analyze, optimize, and design the protocol stack for future modems, we must select the channel model appropriately as it impacts the SINR which inturn affects the overall system performance. As a part of our future research, we will study the impact of blockages on the end-to-end system-level metrics using 3GPP SCM and NYUSIM. Furthermore, we will extend the mmWave module to support multiple spatial streams and frequencies above 100 GHz by leveraging NYUSIM to develop a simulation platform for 6G (sub-THz) networks in ns-3.



\end{document}